\documentclass[prx,twocolumn,superscriptaddress,longbibliography]{revtex4-2}
\usepackage{bm}
\usepackage{amsmath, amsfonts}
\usepackage{amssymb}
\usepackage{times}
\usepackage{graphicx}
\usepackage[table]{xcolor}
\definecolor{lightgray}{gray}{0.9}
\usepackage{mathtools}
\usepackage{float}
\usepackage[colorlinks=true,allcolors=blue]{hyperref}
\usepackage[capitalise,nameinlink]{cleveref}
\usepackage{tikz}
\crefname{section}{Sec.}{Figs.}
\graphicspath{{figures/}}
\begin{document}

\title{Capacitance sensing in bilayer graphene with gate reflectometry}

\author{Tamás Kalmár}
\affiliation{Department of Physics, Institute of Physics, Budapest University of Technology and Economics, Műegyetem rkp.\ 3., H-1111 Budapest, Hungary}
\affiliation{MTA-BME Correlated van der Waals Structures Momentum Research Group, Műegyetem rkp. 3., H-1111 Budapest, Hungary}

\author{Gergő Fülöp}
\email{fulop.gergo@ttk.bme.hu}
\affiliation{Department of Physics, Institute of Physics, Budapest University of Technology and Economics, Műegyetem rkp.\ 3., H-1111 Budapest, Hungary}
\affiliation{MTA-BME Superconducting Nanoelectronics Momentum Research Group, Műegyetem rkp.\ 3., H-1111 Budapest, Hungary}

\author{Bálint Szentpéteri}
\affiliation{Department of Physics, Institute of Physics, Budapest University of Technology and Economics, Műegyetem rkp.\ 3., H-1111 Budapest, Hungary}
\affiliation{MTA-BME Correlated van der Waals Structures Momentum Research Group, Műegyetem rkp. 3., H-1111 Budapest, Hungary}

\author{Olivér Kürtössy}
\affiliation{Department of Physics, Institute of Physics, Budapest University of Technology and Economics, Műegyetem rkp.\ 3., H-1111 Budapest, Hungary}
\affiliation{MTA-BME Superconducting Nanoelectronics Momentum Research Group, Műegyetem rkp.\ 3., H-1111 Budapest, Hungary}

\author{Kenji Watanabe}
\affiliation{Research Center for Electronic and Optical Materials, National Institute for Materials Science, 1-1 Namiki, Tsukuba 305-0044, Japan}
\author{Takashi Taniguchi}
\affiliation{Research Center for Materials Nanoarchitectonics, National Institute for Materials Science,  1-1 Namiki, Tsukuba 305-0044, Japan}

\author{Szabolcs Csonka}
\email{csonka.szabolcs@ttk.bme.hu}
\affiliation{Department of Physics, Institute of Physics, Budapest University of Technology and Economics, Műegyetem rkp.\ 3., H-1111 Budapest, Hungary}
\affiliation{MTA-BME Superconducting Nanoelectronics Momentum Research Group, Műegyetem rkp.\ 3., H-1111 Budapest, Hungary}
\affiliation{HUN-REN Centre for Energy Research, Institute of Technical Physics and Materials Science, Konkoly Thege Miklós út 29-33., H-1121 Budapest, Hungary}

\author{Péter Makk}
\affiliation{Department of Physics, Institute of Physics, Budapest University of Technology and Economics, Műegyetem rkp.\ 3., H-1111 Budapest, Hungary}
\affiliation{MTA-BME Correlated van der Waals Structures Momentum Research Group, Műegyetem rkp. 3., H-1111 Budapest, Hungary}

\begin{abstract}
In the search for topological states in bilayer graphene, capacitance measurement methods are widely utilized in bridge type readouts or source-drain reflectometry setups. In this paper we demonstrate an alternative readout method based on gate reflectometry. We probe the capacitance of a BLG flake encapsulated between hBN and WSe$_2$ through a lumped-element LC resonator coupled to the top gate electrode of the heterostructure. The microwave readout reliably tracks the opening of a displacement field-induced band gap in the device. Compared to simultaneous DC transport measurements, the RF signal proves to be less affected by the percolating transport channels caused by sample disorder. Furthermore, we use the gate based readout to identify a small-angle alignment between the hBN and BLG flake. From the fixed frequency microwave readout, we extract the device capacitance change between the insulating and metallic state of the bilayer and utilize a comprehensive capacitance network model to explain its origin. Finally, we propose an advanced, multiplexed double-resonator readout scheme capable of simultaneously measuring top and bottom gate capacitances to extract the layer polarizability of the system. This work establishes gate reflectometry as a highly scalable and powerful methodology for exploring topological phases and internal charge dynamics in van der Waals heterostructures.
\end{abstract}
\maketitle 
\section{Introduction}\label{sec:intro}

Bilayer graphene (BLG)  holds a unique position among two-dimensional materials due to its extraordinary material properties and flexible applications \cite{Geim2007TheGraphene, McCann2013TheGraphene}.
In contrast to single-layer graphene (SLG) \cite{Novoselov2004, CastroNeto2009}, BLG can be turned into a 2D band insulator, by applying an external displacement field perpendicular to the flake \cite{Oostinga2008Gate-inducedDevices,Taychatanapat2010ElectronicFields,Zhang2009DirectGraphene,Ohta2006ControllingGraphene,Szafranek2010ElectricalTemperature,Castro2007BiasedEffect,Oostinga2008Gate-inducedDevices,Li2016Gate-controlledGraphene,Min2007AbBilayers,Icking2022TransportGraphene}. This allows electron confinement, and recently bilayer quantum dots with a high degree of control \cite{Lee2020TunableContacts,Overweg2018a,Banszerus2020ObservationTransport,Dulisch2025Electric-Field-TunableDot,Kurzmann2019,Allen2012Gate-definedGraphene,Goossens2012,Banszerus2020Single-ElectronGraphene} have been realized as potential hosts for promising topological devices \cite{Overweg2018,Kraft2018,Blanda2026IsolatingDevices,Szentpeteri2025IncreasingPressure}.

Advances in fabrication, including graphite gates \cite{Yankowitz2019,Yankowitz2019a,Overweg2018,Icking2022TransportGraphene,Wang2013} and combination with other van der Waals materials \cite{Dean2010, Arora2020,Gerber2025TunableDevices,Dulisch2025Electric-Field-TunableDot,Szentpeteri2025IncreasingPressure,Seiler2024Interaction-DrivenGraphene,Masseroni2024Spin-orbitHeterostructures} or controlling the twist angle between layers~\cite{Ponomarenko2013, Cao2018a, Cao2018b, Serlin900} opened the way towards novel heterostructures and topological phases, including superconductivity \cite{Holleis2025NematicityGraphene,Cao2018a, Zhang2023EnhancedGraphene,Zhang2024Twist-ProgrammableGraphene,Li2024TunableGraphene,Yankowitz2019}, magnetism \cite{Zhou2022IsospinGraphene,Seiler2022QuantumGraphene,Geisenhof2021QuantumGraphene, Serlin900} and correlated insulator phases \cite{Seiler2024Interaction-DrivenGraphene,Seiler2024Layer-selectiveGraphene,Weitz2010}. The presence of topological states lead to the formation of conducting edge states in the sample, while the bulk of the material remains insulating \cite{Li2016Gate-controlledGraphene,Zhu2017EdgeGraphene}. To distinguish edge- and bulk-related transport effects in standard DC measurements is a non-trivial task \cite{Zhu2017EdgeGraphene,Zeng2019,Veyrat2020HelicalSrTiO3,Young2014TunableState}.

Recent experiments have used capacitance measurements to investigate the electronic structure of graphene-based materials \cite{Xia2009MeasurementGraphene,Droscher2010QuantumGraphene,Yu2013InteractionCapacitance,Young2011CapacitanceProperties,Young2012ElectronicGraphene,Maji2024SuperconductingMaterials,An2026On-chipGraphene,delaBarrera2022CascadeField,Johmen2023Radio-FrequencyBack-Gates,Ranjan2017ContactlessJunctions}. In these measurements, graphene behaves as a semiconducting capacitor plate with the other plate being a metallic gate electrode. In these systems due to the low density of states, adding extra charges also result in an increase of the chemical potential $\mu$. From an electrostatic point of view, this additional energy cost can be incorporated in the total device capacitance as a quantum capacitance term in series with the geometric capacitance \cite{Droscher2010QuantumGraphene,Yu2013InteractionCapacitance} that reduces the total capacitance between the electrodes $C_\mathrm{tot}^{-1}=C_0^{-1}+C_Q^{-1}$. In simple cases, for example in SLG, quantum capacitance is defined as $C_Q=A\nu$, where $A$ is the flake area and $\nu=e^2\partial n/\partial\mu$ is the electronic compressibility in the semi-metal electrode defined by the derivative of carrier density ($n$) with respect to the chemical potential of the layer \cite{Young2012ElectronicGraphene}.

Capacitance sensing in graphene heterostructures has been demonstrated by incorporating the gate electrode of the device into a capacitance bridge circuit \cite{Henriksen2010MeasurementGraphene,Young2012ElectronicGraphene,Hunt2017DirectGraphene,delaBarrera2022CascadeField} and probing the capacitance difference in the arms of the bridge with a few-hundred-kHz signal. The method is effective in recovering the quantum capacitance of the heterostructures, even capable of addressing layer-specific capacitances. However, this technique requires delicate bridge circuits and dedicated sensing transistors, which are difficult to incorporate in compact on-chip solutions in highly scalable designs. A novel approach is to utilize microwave methods to measure the capacitance in BLG \cite{Maji2024SuperconductingMaterials,Vigneau2023ProbingReflectometry,Johmen2023Radio-FrequencyBack-Gates,Ruckriegel2024ElectricResonator,Hecker2025Radio-FrequencyDots}. In these setups the sample is incorporated in a resonator and capacitance change is determined from the resonance frequency shift in the reflected RF signal. These techniques extend the bandwidth of capacitance measurements and offer better scalability with on-chip resonator designs \cite{Maji2024SuperconductingMaterials,An2026On-chipGraphene}. So far, reflectometry studies of BLG were mainly focused on setups where the readout signal is coupled to the flake via an ohmic contact \cite{Johmen2023Radio-FrequencyBack-Gates,An2026On-chipGraphene}. The weakness of this type of readout -- source-drain reflectometry -- comes from its inability to probe a specific layer.

In gate reflectometry \cite{Maji2024SuperconductingMaterials}, the microwave signal is coupled to a gate electrode and thus a capacitive coupling is established between the device and the resonator. This method proved to be sensitive to internal charge dynamics in non-contacted SLG flakes as well \cite{Ranjan2017ContactlessJunctions}, allowing for non-invasive characterization that is unhindered by the presence of the contact electrodes.

In this paper, we probe the capacitance of a dual-gated BLG device by measuring the reflected microwave signal of a resonator coupled to the top gate electrode. We use the gate-reflectometry results to reliably track the external displacement-field-induced band gap of the device and identify a small-angle moiré alignment between the BLG and a hBN flake in the heterostructure. Furthermore, we also extract the device capacitance from the measurements, which we explain quantitatively with the comprehensive capacitance matrix model of the system consisting of geometric and solid-state terms. Finally, we present the principles of a dual-resonator gate reflectometry readout scheme that can be used to probe asymmetry between the layers of BLG. Our work develops the methodology for comparing microwave results with band structure measurements, opening an opportunity to use this technique for the analysis of van der Waals heterostructures.

\section{Sample and measurement setup}
\begin{figure}[h]
\includegraphics{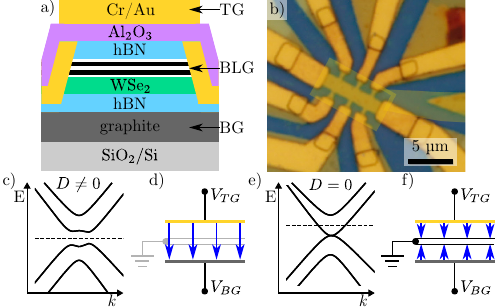}
\caption{\label{fig:fig1} a) hBN/BLG/WSe$_2$/hBN heterostructure with Cr/Au edge contacts and top gate electrode and graphite back gate on SiO$_2$/Si substrate. b) Optical image of the sample etched in a Hall-bar geometry before the deposition of the top gate electrode (semitransparent green area). c) Band structure of BLG around a $K$-point in the gap insulator phase. The Fermi level is marked with the black, dashed line, positioned in the middle of the gap. d) Capacitor schematic of BLG in the insulator phase, with the electric field penetrating through the BLG. e) Band structure of BLG in the metallic phase, with the Fermi level lying at finite energy. f) Capacitor schematic for BLG in the state of panel e. The high charge carrier density in the graphene layers screens the electric field, preventing the penetration.}
\end{figure}

We investigate a BLG flake encapsulated between WSe$_2$ and hBN in a Hall-bar geometry, as shown in Fig.\,\ref{fig:fig1}a. Below the heterostructure a graphite layer serves as the backgate electrode and the metallic top gate electrode is placed on top of it with Al$_2$O$_3$ dielectric. We use an undoped Si/SiO$_2$ substrate to minimize on-chip parasitic capacitances. The optical microscope image of the device is shown in Fig.\,\ref{fig:fig1}b.
With two gate electrodes we can achieve independent control of the charge carrier density and the external displacement field perpendicular to the sample. The band structures at finite and zero displacement field are shown in Fig.\,\ref{fig:fig1}c and Fig.\,\ref{fig:fig1}e, respectively. In the insulator phase the displacement field penetrates through the BLG (Fig.\,\ref{fig:fig1}d), in contrast to the metallic phase, where due to the charge carriers the displacement field will be partially screened \cite{Maji2024SuperconductingMaterials} (Fig.\,\ref{fig:fig1}f). In the two cases we expect different effective capacitances between the gate electrodes and the BLG.

\begin{figure}[h]
\includegraphics{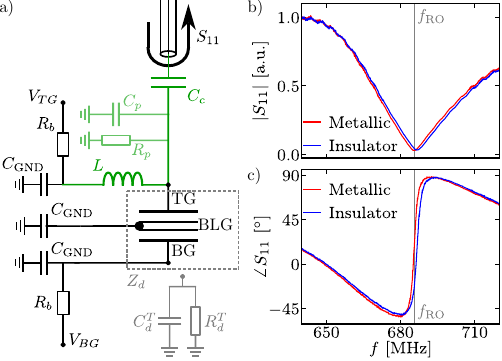}
\caption{\label{fig:fig2} a) Circuit diagram of the reflectometry setup. A meander-shaped NbTiN inductor ($L=61$\,nH) was connected to the top gate of the device, shunting it to ground through a capacitor $C_{\rm{GND}}$. We used an SMD capacitor ($C_c=0.5$\,pF) to couple the resonant circuit with the device to the readout line. Parasitic capacitance and ohmic losses are represented in $C_p$ and $R_p$. Using $C_{\rm{GND}}=1$\,nF and $R_b=250$\,k$\Omega$ we established RF grounds and bias tees on the sample holder PCB, in the close vicinity of the device chip with $Z_d$ impedance (grey rectangle). The RF equivalent $Z_d$ impedance of the device is shown with the grey schematic in the bottom right, formed by a the total top gate capacitance and loss ($C_d^T,R_d^T$) b,c) Magnitude and phase of the resonator response around resonance $f_0\approx686$\,MHz in two different gate configurations exhibit a frequency shift $\delta f_0\approx1$\,MHz. The red curve belongs to metallic phase of the BLG at $V_{BG}=-0.45$\,V, $V_{TG}=-1.39$\,V and DC resistance $R_\mathrm{4P}\approx2.7~\mathrm{k}\Omega$, while the blue shows RF response in the band insulator phase at $V_{BG}=8.11$\,V, $V_{TG}=-8.23$\,V, and $R_\mathrm{4P}\approx650~\mathrm{k}\Omega$. Grey lines indicate the $f_\mathrm{RO}$ readout frequency used during the fixed frequency measurements.}
\end{figure}

Figure \ref{fig:fig2}a shows the schematic of the measurement setup. The top gate electrode of the bilayer sample is connected to a lumped element LC circuit, then coupled to a $Z_0=50$\,$\Omega$ transmission line. In the readout resonator we use a modular architecture, we wire bond a superconducting NbTiN meander fabricated on a separate chip in parallel with the BLG device. In an independent measurement, we determine the inductance of the meander to be $L=61$\,nH.
Additionally, we use an SMD capacitor $C_c=0.5$\,pF to connect the tank circuit to the readout line. This coupling capacitor reduces the RF loss coming from the 50\,$\Omega$ electrical environment \cite{Ahmed2018Radio-FrequencySensing}. In the circuit schematic, parasitic capacitance -- arising between the conductive traces on the chip and the PCB as well as the bonding wires -- and ohmic loss -- mainly dominated by the loss dielectric materials in the setup -- are incorporated in $C_p$ and $R_p$, parallel with the inductor. The total impedance of the circuit is described by
\begin{equation}\label{eq:Z_tot1}
    Z_\mathrm{tot}=1/i\omega C_c+(1/i\omega L+1/R_p+i\omega C_p+1/Z_d)^{-1},
\end{equation}
where $\omega$ is the readout frequency and $Z_d$ is the device impedance. We model the BLG sample embedded in the resonator as an effective parallel RC circuit $Z_d^{-1}=1/R_d^T+i\omega C_d^T$ \cite{Ahmed2018Radio-FrequencySensing,An2026On-chipGraphene}. The total resonator loss $1/R_\Sigma=1/R_p+1/R_d^T$ and capacitance $C_\Sigma=C_p+C_d^T$ has contributions from both the BLG device and the RF environment. We neglect the effect of the large-value grounding capacitors ($C_\mathrm{GND}=1$\,nF), since in the operating frequency range they exhibit a low impedance. In the low-loss limit the resonant frequency of the tank circuit is 
\begin{equation}\label{eq:f0}
    f_0=\frac{1}{2\pi\sqrt{L\left(C_c+C_\Sigma\right)}}.
\end{equation}
The resonator is designed to have $Z_\mathrm{tot}\approx Z_0$ at $f_0$,  ensuring a deep resonance curve in the reflected signal.
On the PCB hosting the tank circuit with the device, DC lines enable the low-frequency four-terminal resistance measurement of the sample at $277$\,Hz utilizing two lock-in amplifiers. Separation of the RF and DC readout is achieved by careful placement of bias-tees and RF ground points on the PCB.

We measure the reflection coefficient $S_{11}$ with a vector network analyzer in a dilution refrigerator at $30$\,mK. Figure \ref{fig:fig2}b-c demonstrate the resonance curves in the reflected magnitude and phase in two gate configurations of the device. The internal and coupling quality factors \cite{Ahmed2018Radio-FrequencySensing} are $Q_i\approx 15$ and $Q_c\approx 17$ with $R_\Sigma\approx4$\,k$\Omega$, operating slightly in the undercoupled regime, consistent with the shape of the phase curve. The resonance frequency is $f_0\approx686$\,MHz and using Eq.\,\ref{eq:f0} while neglecting $C_d^T$, we estimate $C_p\approx385$\,fF.
We observe a significant $\delta f_0\approx1$\,MHz frequency shift between the metallic (red) and band insulator (blue) phases \cite{Maji2024SuperconductingMaterials} -- with the gate voltages summarized in the figure caption --, identified with the aid of the resistance measurement carried out simultaneously.
We find the resonance frequency increased in the gapped phase of the material, indicating the reduction of the effective capacitance in the resonator load.

\section{Radio-frequency response as function of carrier density and displacement field}

\begin{figure*}[htb!]
\includegraphics{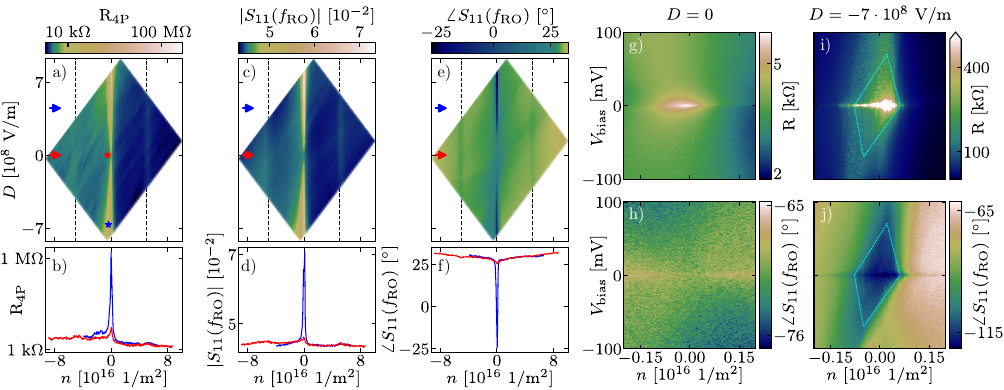}
\caption{\label{fig:fig3}a) Device four-point resistance as a function of displacement field and carrier density. The red and blue stars in the map mark the setpoints that were used in Fig. \ref{fig:fig2}. Around $n=0$ the strong increase in device resistance in the light region of the map indicates the expected charge neutrality in the material. A $\vartheta_m=1^\circ$ alignment between the top layer of the BLG and the hBN results in a secondary, minor resistance increase along along the dashed black lines at $n_m=\pm5\cdot10^{16}$ m$^{-2}$. b) Linecuts from the resistance map at $D=0$ (red) and $D=4.5\cdot10^8$ V/m (blue) displacement values display the tunability of the band gap in the material with external displacement field. c) The magnitude of the reflection coefficient, recorded at the fixed readout frequency also maps the characteristic features of the device transport. At charge neutrality the reflected magnitude rapidly increases, while imprints of the superlattice structure are also present along the same $n_m$ values. d) The linecuts from the reflected magnitude at the same displacement values as panel b display enhanced, tunable RF response around $n=0$, consistent with the band gap in the device. e,f) The reflected phase is effective in capturing the characteristic, tunable transport properties of the device. As the resonance frequency is shifted upwards in the band insulator region, the reflected phase is reduced at the fixed readout frequency. g) Device resistance as a function of DC bias and charge carrier density recorded at $D=0$ reveals minor resistance change in the device. Without external displacement field the charge neutrality region is absent in the material. h) Reflected phase recorded with the measurement in panel g exhibits insignificant changes in the RF response at $D=0$. i)  Device resistance as a function of DC bias and charge carrier density recorded at $D=-7\cdot10^8$\,V/m reveal increased device resistance at finite displacement. A small white region in the middle of the map with $R>500~\text{k}\Omega$ is surrounded by a wide, transition region towards the $R<100~\text{k}\Omega$ resistance region. The cyan lines indicate the boundaries of a bias diamond, corresponding to a band gap of $E_\mathrm{gap}=70$\,meV expected at this displacement field. j) The reflected phase recorded at finite displacement captured a phase decrease in a sharp, diamond-shaped region which aligns well with the theoretically expected diamond size.}
\end{figure*}
Next, we focus on probing the device properties as a function of the displacement field $D$ and carrier density $n$ by tuning the top and bottom gate voltages. For the conversion of gate voltages to $n$ and $D$, we use the geometric capacitances of the top and bottom gates determined in an independent quantum Hall measurement ($C_T^0=10.81$\,fF, $C_B^0=8.93$\,fF, detailed in the Supplementary Material).

Figure \ref{fig:fig3}a shows the DC resistance of the device as a function of displacement field and charge density. A high-resistance region close to $n=0$ is visible, corresponding to the band gap of the BLG. In panel b we compare linecuts from the map along the red arrow, at zero and along the blue one, at finite displacement.
We observe the device resistance to strongly increase along the charge neutrality line, which signals the opening of a band gap in the electronic structure of the device.
Furthermore, along the black dashed lines, around $n_m=\pm5\cdot10^{16}$\,m$^{-2}$ we also observe a minor resistance increase. We attribute this to the formation of a moiré superlattice formed between the hBN and BLG layers \cite{Yankowitz2012EmergenceNitride}.
The small mismatch between the lattice constants and the small rotation between the flakes lead to a renormalization of the band structure at relatively small densities. From the position of the resistance peaks, we obtain a rotation angle of $\vartheta_m=1^\circ$, as detailed in the Supplementary Material.

Simultaneously to the DC readout, we also measure the RF response at a fixed frequency, which we set to be on resonance in the metallic limit at $f_\mathrm{RO}=685.7$\,MHz. The maps in \ref{fig:fig3}c,e show the reflected magnitude and phase as a function of $D$ and $n$. Both quantities exhibit significant response along the charge neutrality line, allowing the opening of the band gap to be clearly traced. Compared to the conducting state, we detect increasing magnitude and decreasing phase response in the band gap. We attribute this to the upward shift of the resonance frequency caused by a net capacitance decrease in the insulator phase. We note that this observation is in agreement with the results of the frequency sweeps shown in Fig.\,\ref{fig:fig2}b.
Linecuts in panels d and f show a sharp RF response in the reflected signal approaching the gap. Both the signal magnitude and phase become more pronounced with larger displacement,  mirroring the behavior observed in the DC readout.
Remarkably, similarly to the DC readout, the RF signal is also sensitive to the superlattice alignment, as shown by the phase and magnitude response at the $n=n_m$ charge density values (dashed black line).

\section{Finite-bias measurements}

To determine the size of the band gap opened by applying the displacement field, we perform bias spectroscopy on the sample. 
In Fig.\,\ref{fig:fig3}g-j the resistance and the fixed-frequency RF response are shown as a function of carrier density and DC bias voltage. Panels g and h display spectroscopy results for the DC resistance and reflected phase recorded at zero displacement. A small increase in the device resistance is visible around charge neutrality, while the reflected phase does not display strong characteristic features. However, at $D=-7\cdot10^{8}$\,V/m in panel i we observe an extended region with a significant resistance increase resembling a diamond shape with blurred edges. Following the theoretical model of McCann and Koshino \cite{McCann2013TheGraphene,Icking2022TransportGraphene} we approximate the gap size at this displacement field to $E_\mathrm{gap}\approx\mathrm{70}$\,meV depicted by the cyan diamond in the map. In the experimental data, we observe a strong increase in the device resistance in this region. From the highly resistive, white region -- with device resistance $R_{4P}>500~\text{k}\Omega$ -- we estimate the gap size to $15$\,meV. We attribute the apparent reduction of the band gap to the effect of a disorder potential present in the measured heterostructure. The disorder potential causes local fluctuations in the Fermi energy \cite{Rossi2011InhomogenousGraphene} and gives rise to tail states with subgap energies \cite{Icking2022TransportGraphene}.
These states can form percolating electron paths  that enable charge transport in the sample, and result in a reduced apparent gap size.

As established by the zero-bias experiment, the low-conducting phase can be identified by the strong negative phase shift in the reflectometry data. Thus, we identify the dark region of panel j as the gapped state of BLG. The data captured by the reflectometry readout exhibits less variation within the insulating region compared to the resistance data of panel i.
The response in the insulating state is more uniform because the RF readout averages the quantum capacitance across the entire gated flake, and it is not selective to conducting channels.

This experiment demonstrates that gate reflectometry provides information complementary to DC transport measurements. While quantum capacitance captures contributions from all electronic states, resistance measurements reflect only those states that produce a net charge current.

\section{Equivalent capacitance network}

In the following, we establish a model for our circuit that allows the extraction of these band-structure-related capacitance corrections for bilayer graphene. We introduce the equivalent capacitance network for the BLG and use exact circuit calculations, rather than relying on the approximate resonance frequency formula Eq.\,\ref{eq:f0}.
For quantitative extraction of the capacitance, we record a high-resolution, fixed-frequency readout of the resonator response, similarly to the measurement in Fig. \ref{fig:fig3}c-f, while tuning the device with the gate voltages in the close proximity of the charge neutrality region. To evaluate the results,  first we calculate the total impedance from the measured reflection coefficient, according to
\begin{equation}\label{ref:eq_zload}
    Z_\mathrm{tot}=Z_0\frac{1+S_{11}}{1-S_{11}},
\end{equation}
where $Z_\mathrm{tot}=Z_\mathrm{tot}^\mathrm{Re}+iZ_\mathrm{tot}^\mathrm{Im}$ includes resonator parameters as well as the device impedance and $Z_0=50$\,$\Omega$ is the transmission line impedance. We extract $R_\Sigma$ and $C_\Sigma$ using the analytical solution of Eq. \ref{eq:Z_tot1}, substituting the known circuit parameters $L,C_c$, detailed in the Supplementary Material.

Fig. \ref{fig:fig4}a shows the extracted total capacitance $C_\Sigma$ as a function of carrier density and displacement field. The capacitance exhibits a similar structure to Fig. \ref{fig:fig3}a,c,e with decreased values in the insulator phase. In Figure \ref{fig:fig4}b the blue linecut from the capacitance map at $D=8.5\cdot10^8$\,V/m reveals a device capacitance decrease of $\Delta C\approx6$\,fF in the gapped region. Using this capacitance value in Eq. $\ref{eq:f0}$ with the circuit parameters would return a resonance frequency shift of around $\delta f_0\approx2.3$\,MHz, which is almost twice as much as the observed shift in Fig. \ref{fig:fig2}b,c. We attribute this difference to the relatively lossy resonant circuit in our setup, which exhibits a resonance frequency sensitive to ohmic losses in the circuit hindering the applicability of Eq. \ref{eq:f0} (detailed in the Supplementary Material).

We note that the extracted resonator loss was $R_\Sigma\approx4~\text{k}\Omega$ with a few-percent change in the gapped state of the device as shown in the Supplementary Material.

\begin{figure}[h]
\includegraphics{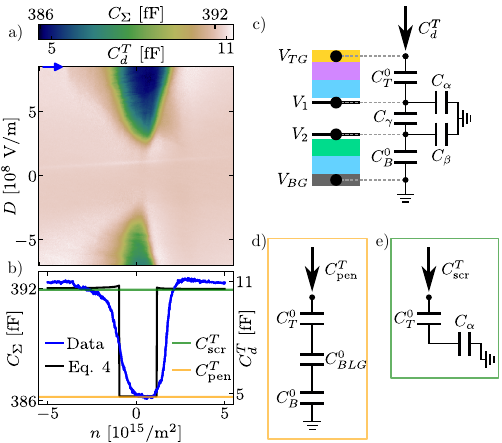}
\caption{\label{fig:fig4} a) Total capacitance ($C_\Sigma$) extracted from the reflectometry signal on the top gate electrode, around charge neutrality.
b) Line cut from panel a at $D=8.5\cdot10^8$\,V/m (blue) is in good agreement with the BLG capacitance model (black) described by Eq. \ref{eq:CTT}, with compressibility values calculated at the same displacement field and over the experimental carrier density range. Orange and green lines are the results of approximating formulas for the penetration (schematic in panel d) and screening capacitance (schematic in panel e), respectively.
c) Effective capacitance network of the BLG device used to describe the observed capacitance. In the heterostructure the top gate, layer 1, layer 2, bottom gate voltage nodes are connected according to the schematic on the right. $C_T^0$, $C_B^0$ are the geometric gate capacitances, while $C_\alpha,C_\beta,C_\gamma$ contain the elements of the electron compressibility matrix of BLG.
d) Approximate device top gate capacitance in the insulator phase, when the device can be modeled with the penetration capacitance $C^T_\mathrm{pen}$.
e) Approximate device top gate capacitance in the metallic phase, when the device can be modeled with the screening capacitance $C^T_\mathrm{scr}$.}
\end{figure}

To explain our findings, we describe the device using an effective lumped-element capacitance network shown in Fig. \ref{fig:fig4}c based on the theoretical results of Ref. \cite{Young2011CapacitanceProperties}. Each layer (from top to bottom: top gate, BLG layer 1, BLG layer 2, bottom gate) of the heterostructure is described by its potential $V_{TG},V_1,V_2$ and $V_{BG}$. The gates are coupled capacitively to the nearest layer, with geometric capacitances $C_T^0$ and $C_B^0$, for the top and bottom gate, respectively.
Layer-specific capacitances ($C_\alpha, C_\beta$) from the BLG layers to the ground and the interlayer capacitance ($C_\gamma$) between the layers contain elements from the electronic compressibility \cite{Yan2011CompressibilityGraphene} tensor of the material, $\hat{\nu}_{ij}=-e^2{\partial n_i}/{\partial V_j}$,
where $n_i$ is the layer-specific charge density. We define the layer-specific capacitances for layer 1 and 2 as $C_\alpha=A_\mathrm{flake}\left(\nu_{11} + \nu_{12}\right)$ and $C_\beta=A_\mathrm{flake}\left(\nu_{22} + \nu_{21}\right)$, where $A_\mathrm{flake}=14.7$\,\textmu m$^2$ is the flake size. The interlayer capacitance $C_\gamma=A_\mathrm{flake}\left(\mathcal{C}_{BLG}^0-\nu_{12}\right)$ contains both geometric and quantum per-area capacitances, where $\mathcal{C}_{BLG}^0=\varepsilon_0/0.3$\,nm, the geometric per-area capacitance of two parallel metallic plates at the BLG interlayer distance and $\nu_{12}=\nu_{21}$ is the off-diagonal element of $\hat{\nu}$.
In the reflectometry readout on the top gate we ensure that the bottom gate electrode is always on an effective RF ground through a small-impedance capacitor. In this case the analytic formula for the device top gate capacitance can be written as
\begin{equation}\label{eq:CTT}
C_d^\mathrm{T}=\left(\left(C^0_B+C_\beta\right)\parallel C_\gamma+C_\alpha\right)\parallel C^0_T,
\end{equation}
using the elements of the network in Fig. \ref{fig:fig4}c and $\parallel$ is the parallel operation. During the measurement any additional device independent parasitic capacitance would be added to $C_d^T$ as on offset. In this case the total measured capacitance is $C_\Sigma=C_d^T+C_p$. We note that according to our model, the BLG device always acts as a capacitive load of the resonant circuit, even in the insulating state.
Therefore, we cannot determine the parasitic capacitance by depleting the BLG and, consequently, we can only determine capacitance changes.

In panel b the black line shows $C_d^\mathrm{T}$ calculated from Eq. \ref{eq:CTT}, where we calculate the elements of $\hat{\nu}$ following Ref.\,\cite{Young2011CapacitanceProperties} with details in the Supplementary Material. Choosing $C_p=381.7$\,fF, the measured dataset (Fig. \ref{fig:fig4}b blue) agrees well with the analytical calculations (Fig. \ref{fig:fig4}b black). We note that our calculation does not account for impurities in the BLG that would cause the capacitance transition to smooth out \cite{Young2011CapacitanceProperties}, as observed in the measurement data.

In the band insulator phase, the capacitance of BLG is dominated by the interlayer geometric capacitance, $C_{BLG}^0\gg C_\alpha,C_\beta$.
In this case majority of the AC current flows through the cross section of the sample, this is the penetration limit depicted in Fig. \ref{fig:fig1}c,d. Fig. \ref{fig:fig4}d represents the equivalent capacitance network with total penetration capacitance $C_\mathrm{pen}^T\approx(1/C_T^0+1/C_{BLG}^0+1/C_B^0)^{-1}$. 
The orange solid line in panel b shows the value of the penetration capacitance calculated from the geometric capacitances of the investigated device, which approximates the value of Eq. \ref{eq:CTT} (panel b, black) in the band gap. We note that the value of the penetration capacitance is equal from both gate electrodes.

In the metallic case the network is dominated by the layer-specific capacitance $C_\alpha\gg C_\gamma$, providing a low-impedance route for AC current towards the grounded ohmic contacts of the device. This is the screening case, where the equivalent capacitance model can be described by the network of Fig. \ref{fig:fig4}e. Using the calculated compressibility of BLG, we also calculate the top gate screening $C_\mathrm{scr}^T=(1/C_T^0+1/C_\alpha)^{-1}\approx C_T^0$ (panel b, green) and find that its value is consistent with Eq. \ref{eq:CTT} in the metallic limit. We note that the expression of the screening capacitance is analogous to the quantum capacitance in single layer graphene, where the geometric gate capacitance is in series with the quantum capacitance of SLG.

\section{Layer polarizability measurement through multiplexed microwave readout}

\begin{figure}[h]
\includegraphics{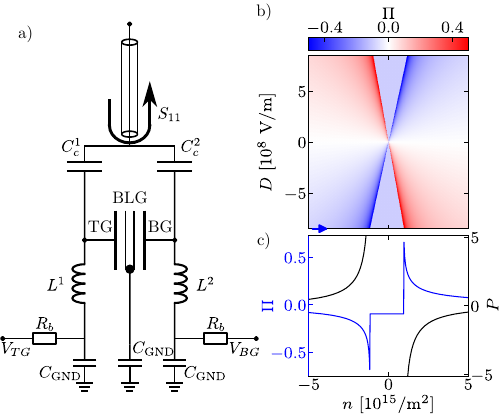}
\caption{\label{fig:fig5}a) Circuit diagram of a double-resonator capacitance readout circuit. Both resonators are coupled to the same readout line via separate coupling capacitors ($C_c^1,C_c^2$). The inductors ($L^1,L^2$) in the resonators are in parallel with the gate electrodes of the capacitively grounded BLG flake. To keep the schematic simple, we do not mark explicitly the parasitic $R_p^{1,2}$ and $C_p^{1,2}$ elements in parallel with the inductors. Individual gating of the electrodes can be established through the inductors using bias tees with high-value resistors to isolate the DC circuity and low-impedance capacitors to establish RF ground nodes. b) The $\Pi$ polarizability as a function of displacement field and charge density calculated from Eq. \ref{eq:Pitilde}, with the geometric capacitance values taken from the experiment. For the BLG capacitance we calculated the elements of the compressibility matrix. Red regions in the map indicate higher compressibility on layer 1. Blue regions are related to the same effect, but on layer 2. In the charge neutrality region $\Pi\approx0$. c) Comparison of $\Pi$ (blue,left axis) and $P$ (black, right axis), shows that both quantities can track internal charge imbalance. In the region, where $P>0$ indicating a surplus in occupied charge states on layer 1, $\Pi<0$, indicating that layer 2 is more compressible, due to lower carrier density on this layer.
}
\end{figure}

In this section we propose a natural extension of the single-resonator measurement, a dual-gate reflectometry setup, shown in Fig. \ref{fig:fig5}a, where two resonators are connected to the gates of the device in the same circuit topology.
If the two resonances are well-separated compared to their broadenings and the arising frequency shifts, the reflected signal can be probed in a multiplexed scheme and the device top ($C_d^\mathrm{T}$) and bottom gate capacitance ($C_d^\mathrm{B}$) can be measured simultaneously. Here $C_d^\mathrm{B}$ can be written using the capacitor network of Fig. \ref{fig:fig4}c, and takes a similar form as Eq. \ref{eq:CTT}:
\begin{equation}\label{CBB}
C_d^B=\left(\left(C^0_T+C_\alpha\right)\parallel C_\gamma+C_\beta\right)\parallel C^0_B.
\end{equation}
While $C_d^T$ and $C_d^B$ alone can be used to investigate the gap of bilayer graphene, measuring both quantities at the same time allows us to probe the asymmetry between the layers.

To recover the polarizability from the multiplexed capacitance readout on the gate electrodes it is effective to use the geometric-capacitance-stripped capacitances, $K_T=(1/C_d^T-1/C_T^0)^{-1}\approx(C_\beta\parallel C_\gamma)+C_\alpha$ and $K_B=(1/C_d^B-1/C_B^0)^{-1}\approx(C_\alpha\parallel C_\gamma)+C_\beta$. Using $K_T$ and $K_B$, the layer polarizability can be defined:
\begin{equation}\label{eq:Pitilde}
    \Pi=\frac{K_T-K_B}{K_T+K_B}\approx\frac{\nu_{11}-\nu_{22}}{2 C_{BLG}^0+\nu_{11}+\nu_{22}}=\frac{C_\alpha-C_\beta}{C_\alpha+C_\beta+2C_\gamma},
\end{equation}
which describes the asymmetry in the layer specific capacitances in the BLG. We note that the quantity is analogous to the quantity defined in Ref. \cite{Young2012ElectronicGraphene} and able to probe the charge-flavor response \cite{delaBarrera2022CascadeField,Hunt2017DirectGraphene,Young2012ElectronicGraphene} through the difference of the diagonal elements of the compressibility matrix $\hat{\nu}$.

Figure \ref{fig:fig5}b shows the calculated value for $\Pi$ as function of displacement and carrier density, using the capacitance values of the previous section. In the $\Pi>0$ case the layer compressibility is higher on layer 1, so on this layer there are more unoccupied charge states compared to layer 2 and vice versa. Near the gap edge, the interlayer imbalance manifests itself as a strong signal in the polarizability. While for large carrier densities $\Pi=0$. Note that the polarizability switches sign with the displacement field. This behavior originates from the fact that in BLG under a nonzero D-field, the wavefunctions are centered on opposite layers for the conduction and valence bands; this layer polarization completely reverses when the D-field is inverted.

In Figure \ref{fig:fig5}c we compare the polarizability (blue, left y-axis) with the charge polarization (black, right y-axis), defined as
\begin{equation}
    P=\frac{n_1-n_2}{n_1+n_2},
\end{equation}
where $n_1$ and $n_2$ are the layer-specific charge densities. We use the values calculated along the $D=-9\cdot10^8$\,V/m line (blue arrow in panel b). The calculations show that outside the gap the polarizability always assumes a value that is to reduce the polarization in the system, $\Pi\propto -P$.

The proposed multiplexed readout method opens a way towards direct charge-flavor measurements through the polarizability of the system.

\section{Conclusion}
In this work, we utilized microwave gate reflectometry to investigate the capacitive properties of a dual-gated BLG flake through the top gate electrode. We find the reflectometry signal effective at identifying the band insulator phase of the material as a function of carrier density and external displacement field. Furthermore, the size of the band gap captured in the RF readout is less impaired by the effects of the percolating conductance channels present in the device due to disorder than that of a standard DC readout. This indicates that the measurement method can be a favorable way to investigate the bulk properties in topological platforms.

In the measured sample a small-angle alignment between the top graphene layer and the hBN dielectric above it results in a secondary Dirac-point in the electronic spectrum. The reflectometry results are in agreement with independent quantum Hall measurements returning the twist angle of the flakes demonstrating the effectiveness of the method for future twistronic applications.

With combined solid-state and radio-frequency calculations we extracted the device capacitance and identified the underlying capacitance network that can be used to model the RF behavior of BLG samples. Our results confirm that in the insulator phase the electric field penetrates through the device, while in the metallic limit the the interlayer capacitance is less dominant, as the individual layers provide a low-impedance route towards the ground for the AC current. We extract $6$\,fF capacitance change in the two limits, which is in agreement with our theoretical calculations considering the electron compressibility in the system.

Furthermore, supported by additional calculations, we propose a dual-gate reflectometry method with multiplexed readout for the determination of layer polarizability. We note that with the usage of on-chip resonator the method can be an effective, scalable option to probe layer imbalance in graphene heterostructures.

The presented gate reflectometry study   brings together the measurement method that already proved its viability in nanowires, quantum dots and spin qubits with the flagship platform of 2D topological materials, opening new ways in the investigation of topological states in van der Waals heterostructures.

\section*{Acknowledgments}
This research was supported by the Ministry of Culture and Innovation and the National Research, Development and Innovation Office within the Quantum Information National Laboratory of Hungary (Grant No.\ 2022-2.1.1-NL-2022-00004). We acknowledge funding from the 2DSPIN-TECH Flagship project, the European Research Council ERC project Twistrain and COST Action CA 21144 superQUMAP. This research was supported by the Bolyai János Research Scholarship of the Hungarian Academy of Sciences. Supported by the EKÖP-25-4-II-BME-46 University Research Scholarship Program of the Ministry for Culture and Innovation from the source of the National Research, Development and Innovation Fund. K.W. and T.T. acknowledge support from the CREST (JPMJCR24A5), JST and World Premier International Research Center Initiative (WPI), MEXT, Japan.

\section*{Author Contributions}
T. K., G. F. and B. Sz. performed the experimental studies. T. K., G. F. and B. Sz. carried out the theoretical simulations and data analysis. B. Sz., O. K., K. W. and T. T. synthesized the materials and fabricated the devices. G. F., Sz. Cs. and P. M. supervised the work. The manuscript was written by T. K and G. F. with input from all other coauthors.

\section*{Competing Interests}
The authors declare no competing interests.

\section*{Data Availability}
The data that support the findings of this study are available from the corresponding author upon request.
\bibliography{references}

\end{document}


\title{Supplementary Material for ``Capacitance sensing in bilayer graphene with gate reflectometry''}

\author{Tamás Kalmár}
\affiliation{Department of Physics, Institute of Physics, Budapest University of Technology and Economics, Műegyetem rkp.\ 3., H-1111 Budapest, Hungary}
\affiliation{MTA-BME Correlated van der Waals Structures Momentum Research Group, Műegyetem rkp. 3., H-1111 Budapest, Hungary}

\author{Gergő Fülöp}
\email{fulop.gergo@ttk.bme.hu}
\affiliation{Department of Physics, Institute of Physics, Budapest University of Technology and Economics, Műegyetem rkp.\ 3., H-1111 Budapest, Hungary}
\affiliation{MTA-BME Superconducting Nanoelectronics Momentum Research Group, Műegyetem rkp.\ 3., H-1111 Budapest, Hungary}

\author{Bálint Szentpéteri}
\affiliation{Department of Physics, Institute of Physics, Budapest University of Technology and Economics, Műegyetem rkp.\ 3., H-1111 Budapest, Hungary}
\affiliation{MTA-BME Correlated van der Waals Structures Momentum Research Group, Műegyetem rkp. 3., H-1111 Budapest, Hungary}

\author{Olivér Kürtössy}
\affiliation{Department of Physics, Institute of Physics, Budapest University of Technology and Economics, Műegyetem rkp.\ 3., H-1111 Budapest, Hungary}
\affiliation{MTA-BME Superconducting Nanoelectronics Momentum Research Group, Műegyetem rkp.\ 3., H-1111 Budapest, Hungary}

\author{Kenji Watanabe}
\affiliation{Research Center for Electronic and Optical Materials, National Institute for Materials Science, 1-1 Namiki, Tsukuba 305-0044, Japan}
\author{Takashi Taniguchi}
\affiliation{Research Center for Materials Nanoarchitectonics, National Institute for Materials Science,  1-1 Namiki, Tsukuba 305-0044, Japan}

\author{Szabolcs Csonka}
\email{csonka.szabolcs@ttk.bme.hu}
\affiliation{Department of Physics, Institute of Physics, Budapest University of Technology and Economics, Műegyetem rkp.\ 3., H-1111 Budapest, Hungary}
\affiliation{MTA-BME Superconducting Nanoelectronics Momentum Research Group, Műegyetem rkp.\ 3., H-1111 Budapest, Hungary}
\affiliation{HUN-REN Centre for Energy Research, Institute of Technical Physics and Materials Science, Konkoly Thege Miklós út 29-33., H-1121 Budapest, Hungary}

\author{Péter Makk}
\affiliation{Department of Physics, Institute of Physics, Budapest University of Technology and Economics, Műegyetem rkp.\ 3., H-1111 Budapest, Hungary}
\affiliation{MTA-BME Correlated van der Waals Structures Momentum Research Group, Műegyetem rkp. 3., H-1111 Budapest, Hungary}

\maketitle 

\subsection*{Device Fabrication}
\subsubsection*{WSe$_2$/BLG/hBN device}
The van der Waals heterostructure is fabricated using the dry-transfer technique \cite{Kim2016} under ambient conditions. The top hBN crystal is used to pick up the BLG which is followed by the WSe$_2$, the bottom hBN and finally the bottom graphite, which serves as a global bottom gate. After the stack is assembled, the device is fabricated using electron beam lithography techniques to define the different layers. Side contacts are created with reactive ion etching and evaporation of 10\,nm of chromium (Cr) and 80\,nm of gold (Au). After the side contacts are made, the heterostructure is etched in a Hall-bar shape by reactive ion etching using SF$_6$/CHF$_3$/O$_2$ gasses. After this, atomic layer deposition is used to create a 30\,nm aluminium oxide dielectric layer, which isolates the top gate, made of 10/80\,nm Cr/Au. The thickness of top hBN is\,25 nm, the bottom hBN is 38\,nm thick and the thickness of the WSe$_2$ layer is 5\,nm, all measured with atomic force microscopy. The hBN is unintentionally aligned with the graphene with the twist angle of $\vartheta_m\approx1$°.\\
Using the atomic force microscopy layer thickness results with the material typical dielectric constants $\varepsilon_{\text{Al}_2\text{O}_3}=11.5$, $\varepsilon_{\text{hBN}}=4$ and $\varepsilon_{\text{WSe}_2}=6$ the gate capacitance from the gate electrodes towards the bilayer are $\tilde{C}_T=14.7~\text{fF}$ and $\tilde{C}_B=12.6~\text{fF}$. These values overestimate the gate capacitance values extracted from the quantum Hall measurements.
\subsubsection*{NbTiN inductors}

In the reflectometry circuit we use a superconducting inductor to reduce parasitic losses \cite{Ahmed2018Radio-FrequencySensing,Annunziata2010a}. The superconducting inductors are made of high-kinetic-inductance NbTiN on an intrinsic Si wafer with $290$\,nm SiO$_2$. We fabricate a long, meander-shaped structure with standard EBL and sputtering techniques, where the kinetic inductance of the superconducting strip is proportional to the conductor length.
We test the inductor by embedding it in a parallel RLC circuit \cite{Annunziata2010a}. We measure the DC resistance of the circuit, while cooling it down in a He dewar to 4\,K. On the onset of the superconducting transition, the DC resistance drops in the circuit, from this we extract the critical temperature $T_c=14$\,K.
To extract the inductance value, we measure the reflection coefficient of the circuit at $T=4$\,K and fit the resonance curve as a function of frequency with fixed resistance and capacitance values. This measurement yields an inductance of $L=61$\,nH.
\subsection*{Quantum Hall measurements}
\begin{figure}[htpb!]
\centering
\includegraphics[width=.5\linewidth]{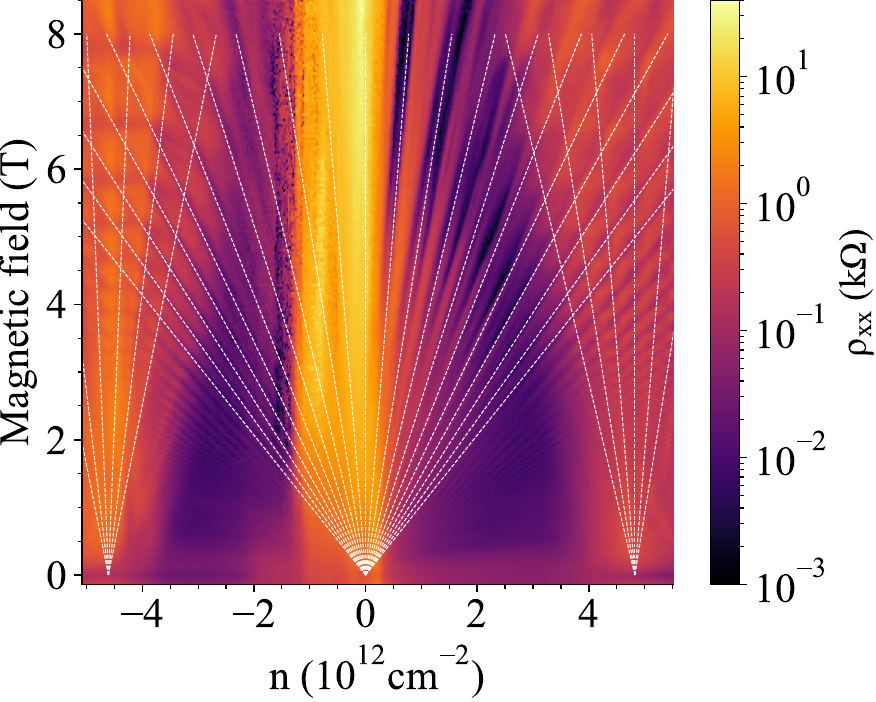}
\caption{\label{fig:supfig_fanplot} Magnetic oscillation measurements of $\rho_{xx}$ at $T=1.4$\,K. The white dashed lines show the Landau levels at different fillings ($\nu=0,\pm4,\pm8\dots$).}
\end{figure}

The charge carrier density can be calculated from the gate voltages as 
\begin{equation}
    n=\alpha_\mathrm{BG}V_\mathrm{BG}+\alpha_\mathrm{TG}V_\mathrm{G}+n_0,
\end{equation}
where $V_\mathrm{BG}$ ($V_\mathrm{TG}$) is the gate voltage of the bottom (top) gate with the lever arm of $\alpha_\mathrm{BG}$ ($\alpha_\mathrm{TG}$) and $n_0$ is the carrier density when the gate voltages are set to zero.
The geometric capacitances are related to the lever arms as
\begin{equation}
    C_{T/B}^0=e A_{\mathrm{flake}} \alpha_\mathrm{TG/BG}.
\end{equation}
To determine the lever arms, geometric capacitances ($C_T^0, C_B^0$) and the twist angle, we have performed Quantum Hall measurements at 1.4 K. We investigated the quantum oscillations in magnetoconductance measurements (Supp. Fig.~\ref{fig:supfig_fanplot}) using the full filling of the Landau levels (LL) being at 
\begin{equation}\label{eq:LLs}
    n=\nu eB/h,
\end{equation}
where $e$ is the elementary charge, $h$ is the Planck's constant, $B$ is the transverse magnetic field and $\nu=0,\pm4,\pm8\dots$ is the filling factor of the LLs. The dark regions in the figure correspond to filled LLs. Eq.\,(\ref{eq:LLs}) was fitted to these regions (white dashed lines) to convert the gate voltages to $n$ thereby yielding the lever arms ($\alpha_\mathrm{BG}=3.79\cdot10^{15}$\,V$^{-1}$m$^{-2}$ and $\alpha_\mathrm{TG}=4.59\cdot10^{15}$\,V$^{-1}$m$^{-2}$).

In Supp. Fig.~\ref{fig:supfig_fanplot}, there are two Landau fans originating from $n_s \approx \pm 4.72 \times 10^{12} \text{ cm}^{-2}$, which is a consequence of the alignment with the hBN crystal. Since $n_s$ corresponds to the full filling of the moiré bands, the superlattice unit cell area is given by $A_s = 4/n_s$, where the factor of 4 accounts for the spin and valley degeneracies. The moiré lattice constant is $\lambda_s = \sqrt{2A_s/\sqrt{3}}$. Taking into account the lattice mismatch $\delta = 1.8\%$ between the graphene and hBN crystals, and the graphene lattice constant ($a = 2.46 \text{ \AA}$), the twist angle $\vartheta_m$ can be extracted using the geometric relation \cite{Yankowitz2012EmergenceNitride}:
\begin{equation}
    \vartheta_m = \arccos \left[ 1 - \frac{1}{2(1+\delta)} \left( \frac{(1+\delta)^2 a^2}{\lambda_s^2} - \delta^2 \right) \right],
\end{equation}
which yields $\vartheta_m \approx 1.01^\circ$.

\subsection*{Measurement setup}

\begin{figure*}[htpb!]
\includegraphics{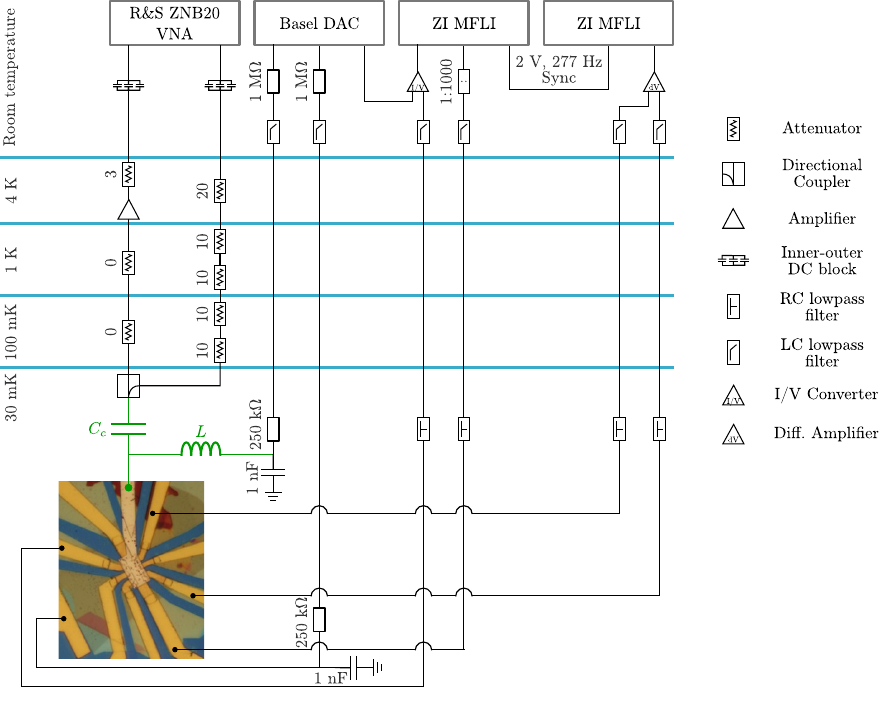}
\caption{\label{fig:setup} Schematic diagram of the cryogenic RF reflectometry measurement setup. The transmission line includes attenuators at each stage of the dilution refrigerator to enable proper thermalization. The resonant circuit coupled to the top gate electrode is marked with green. The four-point resistance measurement was carried out using two lock-in amplifiers, through filtered DC lines.}
\end{figure*}

Supplementary Fig.\,\ref{fig:setup} shows the detailed schematic of the measurement setup that we used to probe the device. The measurements were carried out in the side probe slot of a Leiden Cryogenics dilution refridgerator. The RF signal source and detector was a R\&S ZNB 20 vector network analyzer. From the $4$\,K plate below we had several RF attenuators (2982-5002-dB-CRYO, 2982-6140-dB-CRYO) to ensure proper thermalization of the transmission line. The separation of the incident and reflected signal is realized using a ZX30-12-4-S+ direction coupler. On the $4$\,K plate we used a CITLF2 cryogenic, low-noise amplifier. We equipped both ports of the VNA with inner-outer DC blocks (XMA 2044-7002-18) to avoid ground loops in our setup.
For the supply we used a Basel Precision Instruments (BPI) 8-channel DAC. On the gate lines at room temperature we placed $1$\,M$\Omega$ current-limiting resistors.
The four-terminal resistance measurement was carried out using two Zürich Instruments MFLI lock-in amplifiers, operated at $277$\,Hz readout frequency and synchronized together. One MFLI was responsible for providing the low-frequency voltage bias through a voltage divider, while simultaneously measuring the device current with a BPI Low Noise/High Stability IV converter. The other MFLI measured the voltage drop on the device through a BPI Low Noise/Low Drift Differential Amplifier.
We also included a filter stage on the DC lines at room temperature built out of multiple SMD LC low-pass filters and ferrite beads with an effective cut-off frequency of $\sim100$\,MHz.\\

\begin{figure*}[htpb!]
\includegraphics[scale=1]{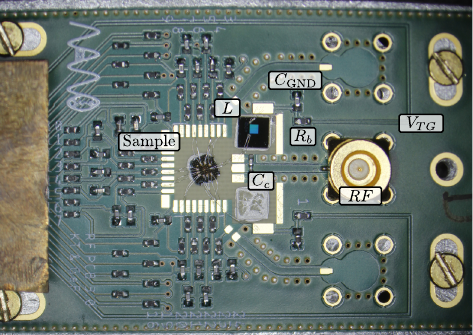}
\caption{\label{fig:pcb} Photograph of the sample holder PCB. The BLG sample was placed in the middle of the board. The RF signal was sent to the sample from the right, through the resonant circuit built from $L$ and $C_c$. The top gate voltage was applied through the inductor using a bias tee, built from SMD components $C_\mathrm{GND}$ and $R_b$.}
\end{figure*}

The BLG sample was placed on a PCB (Supp. Fig.\,\ref{fig:pcb}) thermalized to the MXC plate of the dilution fridge. Apart from the chip the PCB hosted the resonant circuit built from an SMD capacitor (Johanson Technology S201DL) and NbTiN inductor made by our group. To separate the DC and RF signals we had bias-tees present on the card built from $1$\,nF capacitors and $250$\,k$\Omega$ thin-film resistors. On the DC lines without bias-tees we had T-filters present built from two $100$\,$\Omega$ resistors and a $1$\,nF capacitor with a cut-off frequency of $1$\,MHz.

\subsection*{Calculation of the quality factors}
We use the results of Ref.\,\cite{Ahmed2018Radio-FrequencySensing} to calculate the quality factors of the resonances published in the main text. The internal quality factor, which captures the loss in both the device and the resonant circuit, is
\begin{gather}
    Q_i=2\pi f_0(C_c+C_\Sigma)R_\Sigma,
\end{gather}
while the coupling quality factor, determined by the loss in the 50-$\Omega$ electrical environment through the coupling capacitor $C_c$ is
\begin{gather}
        Q_c=\frac{(C_c+C_\Sigma)}{2\pi f_0Z_0C_c^2}.
\end{gather}

\subsection*{Additional bias spectroscopy results}

\begin{figure*}[htpb!]
\includegraphics{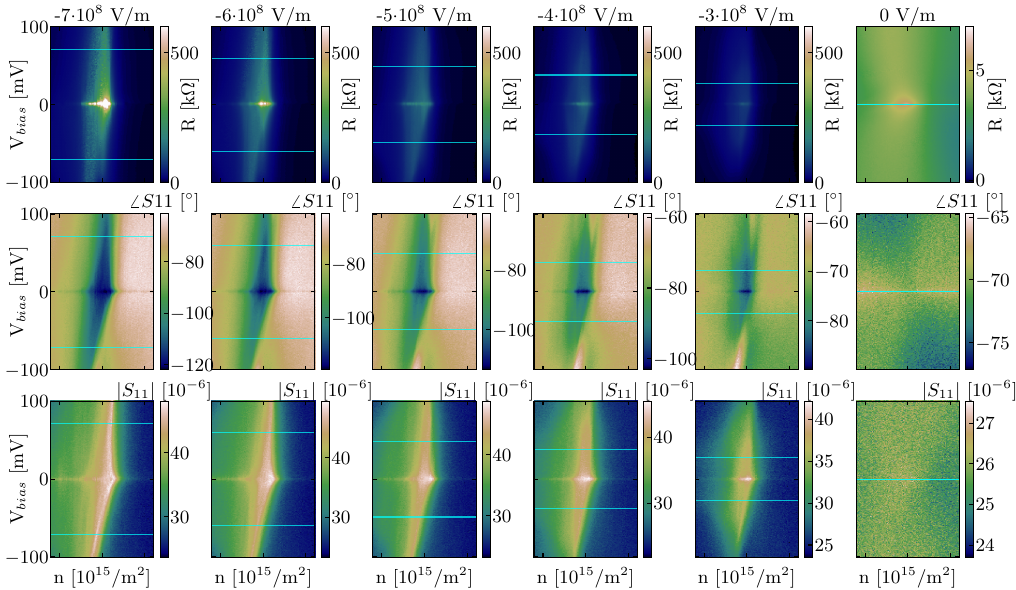}
\caption{\label{fig:diamond} Bias spectroscopy results recorded in the BLG sample. The displacement field is indicated at the top of each column. The first row is the result of the four-terminal DC resistance readout, while the bottom rows are the reflected phase and magnitude respectively. Cyan lines indicate the theoretical value of the gap size at the given displacement field.}
\end{figure*}

In this section we emphasize the applicability of gate reflectometry in bias spectroscopy measurement. The datasets in Fig. 3 of the main text are the part of a measurement set where we record the sample resistance and the reflected signal as a function of carrier density and bias voltage at different displacement fields. In the main text we only publish a dataset at a high value and zero displacement field, however the RF readout proved to be effective at intermediate fields. The visibility of the bias diamonds were higher in the microwave readout and the size of the diamonds aligns well with the theoretical predictions that is indicated with the cyan line in each plot.

\subsection*{Transforming the reflection coefficient to resistance and capacitance values}
In the main text we rely on fixed-frequency results to extract the device capacitance $C_d^T$ and $C_d^B$, in this section we detail the process of the transformation of the reflection coefficient.\\
First, we convert the reflection coefficient to impedance according to
\begin{gather}
S_{11}=\frac{V_{\mathrm{ref}}}{V_{\mathrm{in}}}=\frac{Z_\mathrm{tot}-Z_0}{Z_\mathrm{tot}+Z_0}\label{eq_supp:s11}\\
Z_\mathrm{tot}=Z_0\frac{1+S_{11}}{1-S_{11}}.
\end{gather}

We then assume that the device-related capacitance ($C_d^T$) and effective resistance ($R_d^T$) can be incorporated in a parallel RC circuit -- with contributions from the environment ($C_p,R_p$) -- consisting of $C_\Sigma=C_p+C_d^T,R_\Sigma=1/\left(R_p^{-1}+{R_d^T}^{-1}\right)$, which is in parallel with the inductor in the resonator.

\begin{figure}[htpb!]
\centering
\includegraphics[scale=1]{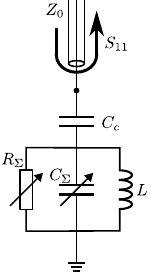}
\caption{\label{fig:resonator} Circuit diagram of the lumped element resonator used in the measurement. The circuits consists of a coupling capacitor ($C_c$), a superconducting inductor ($L$), while the resistance $R_\Sigma$ and capacitance $C_\Sigma$ incorporate the loss and and capacitance contributions coming from the sample as well as parasitic effects.}
\end{figure}

This leads to to the effective picture shown in Supp. Fig. \ref{fig:resonator}, with the total impedance of
\begin{gather}
Z_\mathrm{tot}=\frac{1}{i \omega C}+\left(i \omega C_\Sigma+\frac{1}{R_\Sigma}+\frac{1}{i \omega L}\right)^{-1}\label{Z}.
\end{gather}
The complex impedance of the circuit can be separated to to real and imaginary parts, $Z=Z_r+iZ_i$, using the magnitude and phase of the reflected signal. Both parts can be analytically described with the circuit parametes:
\begin{gather}
Z_r=\left\{R_\Sigma\left[\omega C_\Sigma-\frac{1}{\omega L}\right]^2+\frac{1}{R_\Sigma^2}\right\}^{-1}\label{Zr}\\
Z_i=-\frac{1}{\omega C_c}-\frac{R_\Sigma^2\left(\omega C_\Sigma-\frac{1}{\omega L}\right)}{R_\Sigma^2\left(\omega C_\Sigma-\frac{1}{\omega L}\right)^2+1}\label{Zi}.
\end{gather}
We use Eqs. (\ref{Zr}-\ref{Zi}) to derive the analytical formulas for the values of $C_\Sigma$ and $R_\Sigma$.
\begin{gather}
R_\Sigma=\frac{Z_r^2 C_c^2 \omega^2+Z_i^2 C_c^2 \omega^2+2 Z_i C_c \omega+1}{Z_r C_c^2 \omega^2}\label{eq:R_sigma}\\
C_\Sigma=\frac{Z_r^2 C_c^2 \omega^2+Z_i^2 C_c^2 \omega^2-Z_i C_c^2 L \omega^3+2 Z_i C_c \omega-C_c L \omega^2+1}{L \omega^2\left(Z_r^2 C_c^2 \omega^2+Z_i^2 C_c^2 \omega^2+2 Z_i C_c \omega+1\right)}\label{eq:C_sigma},
\end{gather}
with $L=61$\,nH and $C_c=0.5$\,pF.\\
Then, we subtract the value of the parasitic capacitance $C_p=381.7$\,fF -- that we choose so that the measured capacitance value in the band gap aligns with the penetration capacitance determined by the geometric capacitance values -- from $C_\Sigma$, to arrive at the value of $C_d^T=C_\Sigma-C_p$. This method replaces analytical fits of the resonance curve or the approximation of device capacitance using the resonance frequency shift.\\

\subsection*{Extracted resistance change during the reflectometry readout}
\begin{figure}[htpb!]
\centering
\includegraphics[height=5cm]{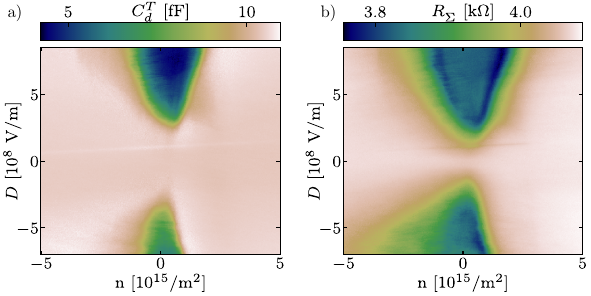}
\caption{\label{fig:RK_result} a) Device top gate capacitance as a function of displacement field and carrier density in the close proximity of the charge neutrality region. b) Resonator loss equivalent resistance simultaneously captured with the capacitance data.}
\end{figure}
Using the fixed-frequency evaluation method, the extracted device capacitance exhibits an approximately $6~\text{fF}$ drop in the insulating phase of the device. This is depicted by the blue regions in Supp. Fig. \ref{fig:RK_result}a as a function of displacement field $D$ and carrier density $n$.
In addition, the extracted resonator loss is also reduced in the gap, shown in panel b, captured simultaneously with panel a. In the metallic phase the RF loss $R_\Sigma\approx4.1~\text{k}\Omega$, while in the band insulator phase $R_\Sigma\approx3.8~\text{k}\Omega$. We attribute this to a gate-dependent RF loss mechanism in BLG \cite{An2026On-chipGraphene}.

\subsection*{The effect of resistance change on the resonance frequency}

In the study we use Eqs. \ref{eq:R_sigma}-\ref{eq:C_sigma} to determine capacitance changes independently from resistive changes, which could not be done by solely relying on resonance frequency shifts and using Eq. 2 of the main text:
\begin{equation}\label{eq:f0}
    f_0=\frac{1}{2\pi\sqrt{L\left(C_c+C_\Sigma\right)}},
\end{equation}
which depends only on the inductive ($L$) and capacitive ($C_p,C_\Sigma$) elements. With a fixed inductance, it suggests that frequency shifts are solely a consequence of capacitive changes. However, as we show below, resistance changes can also lead to changes in the resonance frequency.

\begin{figure}[htpb!]
\centering
\includegraphics[scale=1]{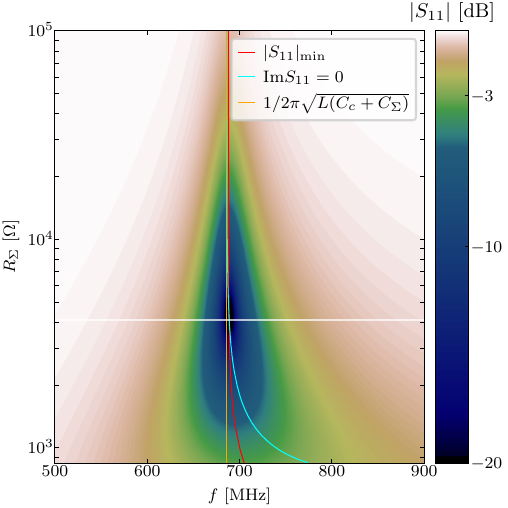}
\caption{\label{fig:f0sim} Simulated reflection coefficient for the readout resonator used in the main text (with the circuit schematic shown in Supp. Fig. \ref{fig:resonator}) as a function of readout frequency at different $R_\Sigma$ values. The orange line shows the resonance frequency, using the approximating formula eq. 2 of the main text. The cyan line shows resonance frequency of the circuit, where by definition the imaginary part of the impedance is zero. The red line indicates the minimum of the resonance curve. Both the red and cyan lines are sensitive to resistance changes around the experimental resistance work point, which is indicated by the white line.}
\end{figure}

Supplementary Figure \ref{fig:f0sim}. shows the simulated reflection coefficient of the lumped element resonator used in the main text (Supp. Fig. \ref{fig:resonator}) for different resistance values in the circuit. The solid white line indicates the resistance value recovered from the experimental data (Supp. Fig. \ref{fig:RK_result}b).

The orange line is the calculated resonance frequency according to Eq. \ref{eq:f0}, which is insensitive to resistive changes. The red line indicates the minimum of the resonance curve as a function of drive frequency, while the cyan line is the analytical calculation for the resonance frequency according to Eq. \ref{Zi}, where the imaginary part of the impedance is zero $Z_i=0$.

According to the simulation the extracted few-hundred-ohm change in the value of $R_\Sigma$ can result in additional resonance frequency shift during the measurement.

Notably, at the low resistance region, the resonance frequency of the resonator (cyan) takes slighly higher values than the minimum of the reflected magnitude (red). Therefore, using the frequency value of the minimum as a resonance frequency for capacitance extraction may hinder the results in this region. 

\subsection*{Capacitance network for bilayer graphene}
\begin{figure}[htpb!]
\centering
\includegraphics[scale=1]{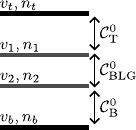}
\caption{\label{fig:BLG_nodes} Schematic explaining for the explanation of the capacitance network of BLG. The parallel plates from top to bottom represent the top gate, BLG layer 1, BLG layer 2 and the bottom gate, with the potential and charge on the plates $v,n$ and the geometric capacitance between the plates $\mathcal{C^0}$.}
\end{figure}

In this section we introduce the background for the equivalent capacitance network that we use to connect the measured capacitance value to the electronic compressibility of the device. We rely on on the results of Ref. \cite{Young2011CapacitanceProperties}. We describe the dual-gated bilayer flake with the network shown in Supp. Fig. \ref{fig:BLG_nodes}. Here the top gate, layer 1 of BLG , layer 2 of BLG and bottom gate nodes are described by their potential ($v_i$) and charge density ($n_i$). The capacitance network of the system connects the potentials and charges in the system:
\begin{equation}
    \mathbf{n}=\mathbf{\hat{\mathcal{C}}}\mathbf{v}.
\end{equation}

In the case of a perfectly metallic system the capacitance matrix only contains geometric capacitance contributions, that satisfy the following equations \cite{Young2011CapacitanceProperties}:
\begin{gather}
\mathcal{C}_{\mathrm{t}}^0\left(v_{\mathrm{t}}-v_1\right)=\frac{1}{2}\left(n_{\mathrm{t}}^0-n_1^0-n_2^0-n_{\mathrm{b}}^0\right), \\
\mathcal{C}_{\mathrm{BLG}}^0\left(v_1-v_2\right)=\frac{1}{2}\left(n_{\mathrm{t}}^0+n_1^0-n_2^0-n_{\mathrm{b}}^0\right), \\
\mathcal{C}_{\mathrm{b}}^0\left(v_2-v_{\mathrm{b}}\right)=\frac{1}{2}\left(n_{\mathrm{t}}^0+n_1^0+n_2^0-n_{\mathrm{b}}^0\right), \\
n_{\mathrm{t}}^0+n_1^0+n_2^0+n_{\mathrm{b}}^0=0,
\end{gather}
with $\mathcal{C}_{\mathrm{t}}^0,~\mathcal{C}_{\mathrm{b}}^0$ and $\mathcal{C}_{\mathrm{BLG}}^0$ the geometry-defined, per-area capacitances in the system. The set of equation can be summarized in a compact matrix equation form $\mathbf{n^0}=\mathbf{\hat{\mathcal{C}}_\mathrm{geom}}\mathbf{v}^0$:
\begin{equation}
\left(\begin{array}{c}
n_{\mathrm{t}}^0 \\
n_1^0 \\
n_2^0 \\
n_{\mathrm{b}}^0
\end{array}\right)=\left(\begin{array}{cccc}
\mathcal{C}_{\mathrm{t}}^0 & -\mathcal{C}_{\mathrm{t}}^0 & 0 & 0 \\
-\mathcal{C}_{\mathrm{t}}^0 & \mathcal{C}_{\mathrm{t}}^0+\mathcal{C}_{\mathrm{BLG}}^0 & -\mathcal{C}_{\mathrm{BLG}}^0 & 0 \\
0 & -\mathcal{C}_{\mathrm{BLG}}^0 & \mathcal{C}_{\mathrm{BLG}}^0+\mathcal{C}_{\mathrm{b}}^0 & -\mathcal{C}_{\mathrm{b}}^0 \\
0 & 0 & -\mathcal{C}_{\mathrm{b}}^0 & \mathcal{C}_{\mathrm{b}}^0
\end{array}\right)\left(\begin{array}{c}
v_{\mathrm{t}}^0 \\
v_1^0 \\
v_2^0 \\
v_{\mathrm{b}}^0
\end{array}\right)
\end{equation}

Due to the finite density of states in graphene, the charge density induced in the layer will be somewhat different than the charge expected from only geometric capacitance contributions ($\mathbf{n^0}$). To calculate the total charge, solid state contributions need to be accounted for, according to $\mathbf{n}=\mathbf{n_0}+\delta\mathbf{n}$, where $\delta\mathbf{n}$ is the additional charge induced by the chemical potential shift ($\delta v_1,\delta v_2$) in BLG due to field effect, which is not present in the metallic gate electrodes. The quantum correction in the charges and potentials of the layers can be summarized as $\delta\mathbf{n}=\hat{\mathbf{\nu}}\delta\mathbf{v}$:
\begin{equation}
\left(\begin{array}{c}
0 \\
\delta n_1 \\
\delta n_2 \\
0
\end{array}\right)=\left(\begin{array}{cccc}
0 & 0 & 0 & 0 \\
0 & \nu_{11} & \nu_{12} & 0 \\
0 & \nu_{21} & \nu_{22} & 0 \\
0 & 0 & 0 & 0 \\
\end{array}\right)\left(\begin{array}{c}
0 \\
\delta v_1 \\
\delta v_2 \\
0
\end{array}\right),
\end{equation}
with $\hat{\mathbf{\nu}}=e^2\delta n/\delta v$ the compressibility matrix of BLG. The elements of the matrix can be given using the symmetrized layer potentials $v_{ \pm}=\frac{1}{2}\left(v_1 \pm v_2\right)$:
\begin{gather}
\nu_{11}=\frac{1}{2} \nu_0 \frac{\left|v_{+}\right|-v_{-} \operatorname{sgn} v_{+}}{\sqrt{v_{+}^2-v_{-}^2}}+\frac{1}{4} \tilde{\nu}_0\label{eq:nu11_def}\\
\nu_{22}=\frac{1}{2} \nu_0 \frac{\left|v_{+}\right|+v_{-} \operatorname{sgn} v_{+}}{\sqrt{v_{+}^2-v_{-}^2}}+\frac{1}{4} \tilde{\nu}_0\label{eq:nu22_def}\\
\nu_{12}=\nu_{21}=-\frac{1}{4} \tilde{\nu}_0,\label{eq:nu12_def}
\end{gather}
where
\begin{equation}
\begin{aligned}
& \tilde{\nu}_0=\nu_0 \ln \left(\frac{2 \Lambda}{e\left(\left|v_{+}\right|+\sqrt{v_{+}^2-v_{-}^2}\right)}\right)
\\
&
\nu_0=2 m e^2 /\left(\pi \hbar^2\right)
\end{aligned}
\end{equation}
are scaling factors for the compressibility in units of F/m$^2$ and $\tilde{\nu}_0$ accounts for ultraviolet cutoff.\\
The charge and layer potential dependence accounting for geometric and quantum corrections in the system can be given as $\mathbf{n}=\mathbf{n^0+\delta n}=\mathcal{\hat{C}}\mathbf{v}$, with the capacitance matrix

\begin{equation}
\mathcal{\hat{C}}=\mathcal{\hat{C}_\mathrm{geom}}+\hat{\nu}=\left(\begin{array}{cccc}
\mathcal{C}_{\mathrm{t}}^0 & -\mathcal{C}_{\mathrm{t}}^0 & 0 & 0 \\
-\mathcal{C}_{\mathrm{t}}^0 & \mathcal{C}_{\mathrm{t}}^0+\mathcal{C}_{\mathrm{BLG}}^0+\nu_{11} & -\mathcal{C}_{\mathrm{BLG}}^0+\nu_{12} & 0 \\
0 & -\mathcal{C}_{\mathrm{BLG}}^0+\nu_{12} & \mathcal{C}_{\mathrm{BLG}}^0+\mathcal{C}_{\mathrm{b}}^0+\nu_{22} & -\mathcal{C}_{\mathrm{b}}^0 \\
0 & 0 & -\mathcal{C}_{\mathrm{b}}^0 & \mathcal{C}_{\mathrm{b}}^0
\end{array}\right)=\left(\begin{array}{cccc}
\mathcal{C}_{\mathrm{t}}^0 & -\mathcal{C}_{\mathrm{t}}^0 & 0 & 0 \\
-\mathcal{C}_{\mathrm{t}}^0 & \mathcal{C}_{\mathrm{t}}^0+\mathcal{C}_\alpha+\mathcal{C}_\gamma & -\mathcal{C}_\gamma & 0 \\
0 & -\mathcal{C}_\gamma & \mathcal{C}_{\mathrm{b}}^0+\mathcal{C}_\beta+\mathcal{C}_\gamma & -\mathcal{C}_{\mathrm{b}}^0 \\
0 & 0 & -\mathcal{C}_{\mathrm{b}}^0 & \mathcal{C}_{\mathrm{b}}^0
\end{array}\right),
\end{equation}
with $\mathcal{C_\alpha}=\nu_{11}+\nu_{12},\mathcal{C_\beta}=\nu_{22}+\nu_{12}$ and $\mathcal{C_\gamma}=\mathcal{C}_{\mathrm{BLG}}^0-\nu_{12}$ per-area inter and layer-specific capacitances, consistent with the main text. $\mathcal{C_\alpha}$ and $\mathcal{C_\beta}$ contain the total layer-specific capacitance corrections for layer 1 and 2 respectively, while $\mathcal{C_\gamma}$ describes the interlayer coupling weakened by solid-state corrections. The circuit schematic described by this capacitance matrix is shown in Fig. 4c of the main text.

\subsection*{Calculation of the BLG capacitances}

In this section we detail the calculations used in the main text to get to the capacitance values $C_\alpha,C_\beta$ and $C_\gamma$ as a function of the experimental parameters $n$ and $D$. 

We use the geometric capacitances $C_T^0=10.81$\,fF, $C_B^0=8.93$\,fF, $C_\mathrm{BLG}^0=\mathcal{C}^0_\mathrm{BLG}A_\mathrm{flake}=434$\,fF and flake size $A_\mathrm{flake}=14.7$\,\textmu m$^2$, consistent with the values of the actual measured device.

For the calcutaion we use $n$ and $D$ as input parameters and calculate the top and bottom gate voltages ($V_\mathrm{TG},V_\mathrm{BG}$) at the setpoint according to:
\begin{equation}
    V_\mathrm{TG/BG}=\frac{n}{2\alpha_\mathrm{TG/BG}}+\frac{2\varepsilon_0D}{e},
\end{equation}
with $\alpha_\mathrm{TG/BG}$ the top or bottom gate lever arm and $\varepsilon_0$ the vacuum permittivity. After extracting the gate voltage setpoint we numerically solve the following set of equations for the symmetrized layer potentials $v_\pm=\frac{1}{2}\left(v_1\pm v_2\right)$ \begin{gather}
    \frac{C^0_T+C^0_B}{e}v_++\frac{C^0_T-C^0_B}{e}v_--C^0_TV_{TG}-C^0_BV_{BG}-en_+=0\\
    -\frac{C^0_T-C^0_B}{e}v_+\frac{4C^0_\mathrm{BLG}+C^0_T+C^0_B}{e}v_--C^0_TV_{TG}+C^0_BV_{BG}-en_-=0,
\end{gather}
where $n_\pm=\frac{1}{2}\left(n_1 \pm n_2\right)$ are the symmetrized layer charges, according to Ref. \cite{Young2011CapacitanceProperties}:
\begin{gather}
n_{+}= \begin{cases}-\nu_0 \sqrt{v_{+}^2-v_{-}^2} \operatorname{sgn} v_{+} & \text {(metal), } \\
0 & \text { (insulator) },\end{cases} \\
n_{-}= \begin{cases}-\nu_0 v_{-} \ln \left(\frac{2 \Lambda}{\left|v_{+}\right|+\sqrt{v_{+}^2-v_{-}^2}}\right) & \text { (metal), } \\
-\nu_0 v_{-} \ln \left(\frac{2 \Lambda}{\left|v_{-}\right|}\right) & \text {(insulator). }\end{cases}
\end{gather}
Then we substitute $v_+$ and $v_-$ into Eqs. \ref{eq:nu11_def}-\ref{eq:nu22_def} to recover the full compressibility matrix at given $n,D$. After this using the flake size and the geometric capacitances, we calculate the values of $C_\alpha,C_\beta$ and $C_\gamma$ according to
\begin{gather}
C_\alpha=A_\mathrm{flake}\left(\nu_{11} + \nu_{12}\right)\label{eq:Ca}\\
C_\beta=A_\mathrm{flake}\left(\nu_{22} + \nu_{12}\right)\label{eq:Cb}\\
C_\gamma=A_\mathrm{flake}\left(\mathcal{C}_{BLG}^0-\nu_{12}\right)\label{eq:Cg}
\end{gather}.

\begin{figure}[htpb!]
\centering
\includegraphics[scale=1]{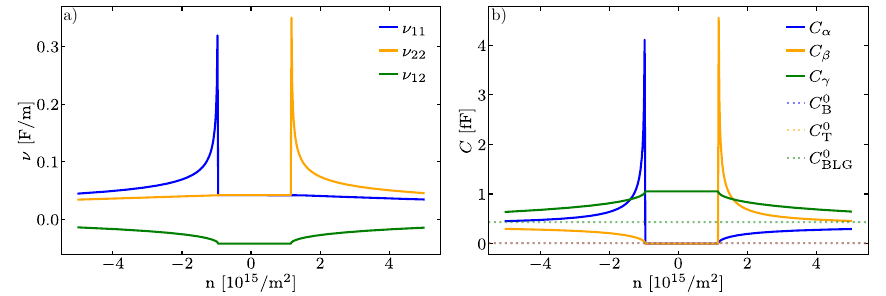}
\caption{\label{fig:BLG_capasim} a) Simulation results for the elements of the compressibility matrix at $D=8.5\cdot10^8$\,V/m as a function of carrier density. b) Simulated values of the capacitance network using the compressibility values in panel a, with the geometric capacitance values $C_T^0=10.81$\,fF, $C_B^0=8.93$\,fF and $C_\mathrm{BLG}^0=\mathcal{C}^0_\mathrm{BLG}A_\mathrm{flake}=434$\,fF. The value of the interlayer and layer specific capacitances are much larger than the geometric gate capacitances, except for the gap, where $C_\alpha=C_\beta=0$.}
\end{figure}

Supp. Fig.\ref{fig:BLG_capasim} shows the values of $\hat{\nu}$ and the layer capacitances as a function of the charge density at $D=8.5\cdot10^8$\,V/m.

\subsection*{Fidelity of RF grounding during the multiplexed readout}
\begin{figure}[htpb!]
\centering
\includegraphics[scale=1]{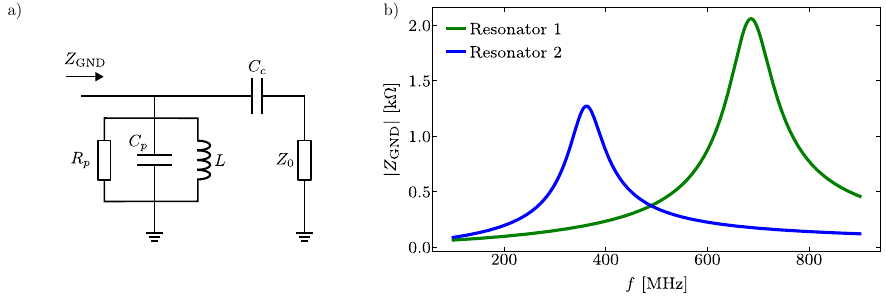}
\caption{\label{fig:multi_impedance} a) Circuit schematic for the definition of the ground impedance $Z_\mathrm{GND}$, which defines the ground impedance seen by the RF current that flows from one gate electrode to the other in the multiplexed setup. b) Calculated ground impedance for Resonator 1 (green) and Resonator 2 (blue) in the multiplexed geometry. On resonance of Resonator 1 (maximum of the green dataset) Resonator 2 provides an RF ground with $Z_\mathrm{GND}\lesssim 200$\,$\Omega$ and vice versa.}
\end{figure}

In case of the multiplexed scheme, we assume that the resonators can be considered independent, and we can apply the method detailed above to the circuit at the two readout frequencies independently. This is justified by their well-separated resonance curves. Furthermore, we note that each resonator provides a low-impedance path to ground for the other one at its respective resonance frequency. In a parallel LC circuit, well below the resonance frequency the inductor exhibits a low impedance, and in the high-frequency limit the capacitor provides the shunt. To calculate the value of the ground impedance exhibited by both resonators for the RF current that passes through between the gate electrodes, we investigate the circuit in Supp. Fig. \ref{fig:multi_impedance}a. We define $Z_\mathrm{GND}=\left(\left(1/R_p+1/i\omega L+i\omega C_p\right)+\left(1/i\omega C_c+Z_0\right)^{-1}\right)^{-1}$, where circuit values are substituted from Supp. Table \ref{tab:rlc_params}. Maximas in the simulated datasets indicate the resonance frequency of the resonators with an impedance $Z_\mathrm{GND}>1$\,k$\Omega$. However, in both cases, on resonance, the other resonator provides an RF ground node with an impedance of few hundred ohms confirming our assumption.

\begin{table}[htpb!]
\centering
\begin{tabular}{ll||ll}
\multicolumn{2}{c||}{Resonator 1}             & \multicolumn{2}{c}{Resonator 2}                                                 \\ \hline\hline
\multicolumn{1}{l|}{$C_c^1$ (pF)}      & 0.5 & \multicolumn{1}{l|}{$C_c^2$ (pF)}      & 1   \\ \hline
\multicolumn{1}{l|}{$R_p^1$ (k$\Omega$)} & 4   & \multicolumn{1}{l|}{$R_p^2$ (k$\Omega$)} & 2   \\ \hline
\multicolumn{1}{l|}{$C_p^1$ (fF)}        & 382 & \multicolumn{1}{l|}{$C_p^2$ (fF)}        & 780 \\ \cline{1-4}
\multicolumn{1}{l|}{$L^1$ (nH)}        & 61  & \multicolumn{1}{l|}{$L^2$ (nH)}        & 108                           \\
\end{tabular}
\caption{Static circuit parameters used in the radio-frequency simulation.}
\label{tab:rlc_params}
\end{table}

\subsection*{Layer polarizability}
In this section we provide additional information of the derivation of layer polarizability from the device capacitances and demonstrate its connection to the layer polarization in BLG.
In the main text $C_d^T$ and $C_d^B$ represents the top and bottom gate capacitance sensed by the RF signal (see Fig.\,4), with the following definitions:
\begin{gather}
C_d^T=\left(\left(C^0_B+C_\beta\right)\parallel C_\gamma+C_\alpha\right)\parallel C^0_T=\left(\left(\left(\left(C_\mathrm{B}^0+C_\beta\right)^{-1}+C_\gamma^{-1}\right)^{-1}+C_\alpha\right)^{-1}+{C_\mathrm{T}^0}^{-1}\right)^{-1}\\
C_d^B=\left(\left(C^0_T+C_\alpha\right)\parallel C_\gamma+C_\beta\right)\parallel C^0_B=\left(\left(\left(\left(C_\mathrm{T}^0+C_\alpha\right)^{-1}+C_\gamma^{-1}\right)^{-1}+C_\beta\right)^{-1}+{C_\mathrm{B}^0}^{-1}\right)^{-1},
\end{gather}
where $C_\alpha,C_\beta$ and $C_\gamma$ are the BLG compressibility dependent capacitances defined by Eqs. \ref{eq:Ca}, \ref{eq:Cb}, \ref{eq:Cg}.

It is effective to use reduce the a expressions by removing the geometric gate capacitances that can be determined through quantum Hall measurements and use the following reduced quantities:
\begin{gather}
    K_T=\frac{1}{\left(C_d^{T}\right)^{-1}-\left(C_T^0\right)^{-1}}=\left(\left(\left(C_\mathrm{B}^0+C_\beta\right)^{-1}+C_\gamma^{-1}\right)^{-1}+C_\alpha\right)^{-1}\\
    K_B=\frac{1}{\left(C_d^{T}\right)^{-1}-\left(C_B^0\right)^{-1}}=\left(\left(\left(C_\mathrm{T}^0+C_\alpha\right)^{-1}+C_\gamma^{-1}\right)^{-1}+C_\beta\right)^{-1}.
\end{gather}
We substitute for $C_\alpha,C_\beta,C_\gamma$ according to Eqs.(\ref{eq:Ca}-\ref{eq:Cg}) and define the experimental layer polarization
\begin{widetext}
\begin{equation}\label{eq:kmkp}
    \Pi=\frac{K_T-K_B}{K_T+K_B}=\frac{\nu_{11}-\nu_{22}-\frac{\left(C_{B L G}^0-\nu_{12}\right)\left(C_T^0+\nu_{11}+\nu_{12}\right)}{C_T^0+C_{B L G}^0+\nu_{11}}+\frac{\left(C_{B L G}^0-\nu_{12}\right)\left(C_B^0+\nu_{12}+\nu_{22}\right)}{C_B^0+C_{B L G}^0+\nu_{22}}}{\nu_{11}+2 \nu_{12}+\nu_{22}+\frac{\left(C_{B L G}^0-\nu_{12}\right)\left(C_T^0+\nu_{11}+\nu_{12}\right)}{C_T^0+C_{B L G}^0+\nu_{11}}+\frac{\left(C_{B L G}^0-\nu_{12}\right)\left(C_B^0+\nu_{12}+\nu_{22}\right)}{C_B^0+C_{B L G}^0+\nu_{22}}}.
\end{equation}
\end{widetext}
The expression simplifies in the typical case where quantum corrections dominate the network over the geometric gate capacitances with the exception of the interlayer capacitance (Supp. Fig. \ref{fig:BLG_capasim}b), which due to the small spacing in the parallel plate picture cannot be neglected. In this limit,
\begin{equation}\label{eq:limPitilde}
    \lim_{C_T^0,C_B^0\rightarrow0}\Pi=\frac{\nu_{11}-\nu_{22}}{2 C_{B L G}^0+\nu_{11}+\nu_{22}}=\frac{C_\alpha-C_\beta}{C_\alpha+C_\beta+2C_\gamma},
\end{equation}
which gives an opportunity to determine the polarizability through the RF response of the system. We note that due to the approximation for the geometric gate capacitances, the value of $\Pi\neq0$ in the gapped region, when $C_0^T\neq C_0^B$.

\begin{figure}[htpb!]
\centering
\includegraphics[scale=1]{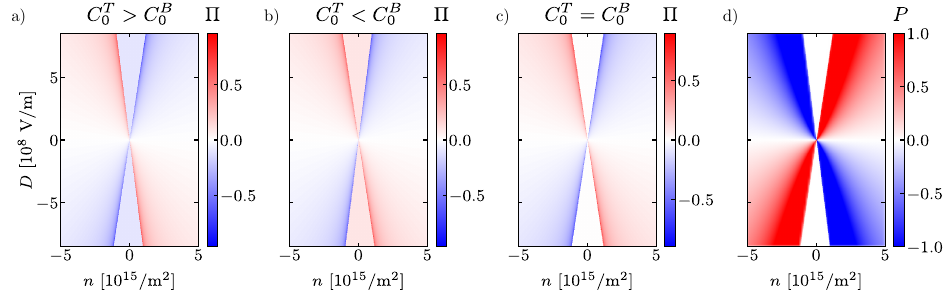}
\caption{\label{fig:polsim} a) Calculated polarizability according to Eq. \ref{eq:kmkp} as function of displacement field and carrier density for $C_0^T=10.8~\text{fF}$, $C_0^B=8.9~\text{fF}$. In the blue region the compressibility or local density of states is higher on layer 2, while in red areas the compressibility of layer 1 is higher. Due to the asymmetry between the geometric gate capacitances, the value of $\Pi$ takes a finite negative value in the gap. b) Calculated experimental layer polarizability according to Eq. \ref{eq:kmkp} for the same $n,D$ range as panel a, but with $C_0^T=8.9~\text{fF}$, $C_0^B=10.8~\text{fF}$. In this case the value of $\Pi$ is a small positive number in the gap, however other features of the map qualitatively match panel a. c) Eq. \ref{eq:kmkp} for the same $n,D$ range as panel a, but with $C_0^T=C_0^B=10.8~\text{fF}$. In this case the value of $\Pi=0$ in the gap. d) Calculated polarization $P=(n_1-n_2)/(n_1+n_2)$ for the same $n$,$D$ range. Blue areas in the map indicate that majority of the charge carriers as located on layer 2, while red areas indicate layer 1 hosts more charge carriers. We note that the positive layer polarization corresponds to negative polarizability.}
\end{figure}

Supplementary Figure \ref{fig:polsim}a,b,c shows the calculated values for polarizability $\Pi$ for different geometric gate capacitance values according to Eq. \ref{eq:kmkp}. We note that the approximation of Eq. \ref{eq:limPitilde} holds only in the case when gate capacitances are symmetric (panel c), in the other cases (panel a,b) $\Pi$ takes a small finite value within the gap. In the blue regions layer 2 has a higher compressibility, while in red areas the effect is reversed. In order to exclude the effect of asymmetric gate capacitances, it is important to fabricate devices where the bottom and the top gate has similar size and only cover the active area of the heterostructure.

Supp. Fig. \ref{fig:polsim}d shows the calculated value for the layer polarization, that is defined by the normalized carrier density difference in the layers $P=\frac{n_-}{n_+}=\frac{n_1-n_2}{n_1+n_2}$, where $n_i$ is the carrier density in layer i.

We note that using the formalism of Ref. \cite{Young2011CapacitanceProperties} the polarizability of this work can be defined as
\begin{equation}
    \Pi=\frac{2\nu_{-+}}{4C^0_{BLG}+\nu_{--}+\nu_{++}}=\frac{2(\nu_{11}-\nu_{22})}{4C^0_{BLG}+2\nu_{11}+2\nu_{22}}=\frac{C_\alpha-C_\beta}{C_\alpha+C_\beta+2C_\gamma},
\end{equation}
with the layer symmetrized charge densities $n_\pm=(n_1\pm n_2)$ and potentials $v_\pm=\frac{1}{2}(v_1\pm v_2)$ and the compressibility matrix as $\nu_{++}=-\partial n_+/\partial v_+=\nu_{11}+\nu_{22}+2\nu_{12}$, $\nu_{--}=-\partial n_-/\partial v_-=\nu_{11}+\nu_{22}-2\nu_{12}$ and $v_{+-}=v_{-+}=-\partial n_-/\partial v_+=\nu_{11}-\nu_{22}$.

\bibliography{references}